\documentclass[]{aastex631}

\newcommand{\hetg}{{\small {\it Chandra}/HETG}}

\newcommand{\gro}{{\small GRO\,J1655-40}}

\graphicspath{{./}{figures/}}

\begin{document}

\title{ {Supernova Archaeology with X-Ray Binary Winds - The Case of GRO J1655-40}} 

\author{Noa Keshet}
\affiliation{Department of Physics, Technion, Haifa 32000, Israel}

\author{Ehud Behar}
\affiliation{Department of Physics, Technion, Haifa 32000, Israel}

\author{Timothy R. Kallman}
\affiliation{NASA Goddard Space Flight Center, Code 662, Greenbelt, MD 20771, USA}









\begin{abstract}

Supernovae are responsible for the elemental enrichment of the galaxy and some are postulated to leave behind a black hole. 
In a stellar binary system the supernova pollutes its companion, and the black hole can accrete back its own debris and emit X-rays. In this sequence of events, which is only poorly understood, winds are ejected, and observed through X-ray absorption lines. Measuring abundances of elements in the wind can lead to inferences about the historical explosion and possibly identify the long-gone progenitor of the compact object. Here, we re-analyze the uniquely rich X-ray spectrum of the 2005 outburst of \gro .
We reconstruct the absorption measure distribution (AMD) of the wind, and find that it increases sharply with ionization from H-like O up to H-like Ca, and then flattens out. The AMD is then used to measure relative abundances of 18 different elements. The present abundances are in partial agreement with a previous work with discrepancies mostly for low-Z elements. The overabundance of odd-Z elements hints at a high-metallicity, high-mass ($\simeq25\,M_\odot$) progenitor.
Interestingly, the abundances are different from those measured in the companion atmosphere, indicating that the wind entrains lingering ambient supernova debris. This can be expected since the current total stellar mass of the binary ($<10\,M_\odot$) is much less than the progenitor mass. 


\end{abstract}



\section{Introduction} \label{sec:intro}

It has long been understood that the source of most elements in the universe are stellar cores that explode in a supernova (SN). 
Only a handful of elements can be measured in the limited visible band observed immediately after SN explosions. 
Measuring abundances in X-ray spectra of SN remnants 
can reveal some elements, hint at the progenitor, and test SN models \citep{Siegel2020, Vink2012}. 
However, a full account of all elements dispersed by the SN has not been obtained from observations.
Here, we indirectly study a historical SN by observing the low-mass X-ray binary \gro\ (also known as Nova Scorpii 1994). \gro\ is an X-ray binary at a distance of 3.2\,kpc with a 6.6\,$M_{\odot}$ black hole (BH) and a low mass 2.7\,$M_{\odot}$ F-star companion in a 2.6\,day period \citep{Casares2014}. 
The implied semi-major axis of the binary is $a=1.1\times10^{12} $ cm.

In 2005, \gro\ went into outburst and the ensuing outflow produced a rich X-ray spectrum measured with the \hetg\ gratings,
which has been the focus of several works \citep{Miller2006, Miller2008, Kallman2009, Fukumura2017, Tomaru2023}. The spectrum features a multitude of blue shifted lines from many elements, and many ions ranging from H-like O to H-like Fe. This likely indicates a range of ionization parameter $\xi = L / (nr^2)$ values, where $L$ is the ionizing luminosity, $n$ the number density, and $r$ the distance from the source.
$\xi$ represents the balance between photo-ionization (flux $\propto L/r^2 $) and recombination ($\propto n$) rates. 
\citet{Miller2008} identified most of the lines in the spectrum, and measured their velocities. Their paper focuses on the absorption lines from Fe meta-stable levels that provide density diagnostics. The high density was used (via $\xi$) to rule out thermal wind driving. Other works suggested that thermal driving might still be viable \citep{Neilsen2012, Tomaru2023}.  
\citet{Kallman2009} performed a global spectral fit of a photo-ionized plasma model. Their best fit model (No. 6) yields an outflow velocity of $-375$\,km\,s$^{-1}$, a total column density of $N_{\rm H} = 10^{24}$\,cm$^{-2}$, and mostly solar elemental abundances using a single ionization component of $\xi = 10^4$\,erg\,cm\,s$^{-1}$.
\citet{Tomaru2023} fitted the \hetg\ outburst spectrum with two photoionized components of $\log\xi = 4.1, 3.4$ (cgs units here and throughout the paper), and with a total column density $N_{\rm H} = 5\times10^{24}$\,cm$^{-2}$.
No global fitted model is statistically satisfactory, due to the exquisite quality of the \hetg\ spectrum.

The present work utilizes a different approach of ion-by-ion column density fitting \citep{Behar2003, Holczer2007, Peretz2018}, complemented by individual line measurements and curve of growth analysis. 
These column densities are used to reconstruct of the ionization distribution of the wind.
 

\section{Observations and Data} 

\begin{figure}
    \centering
    \includegraphics[width=0.32\textwidth]{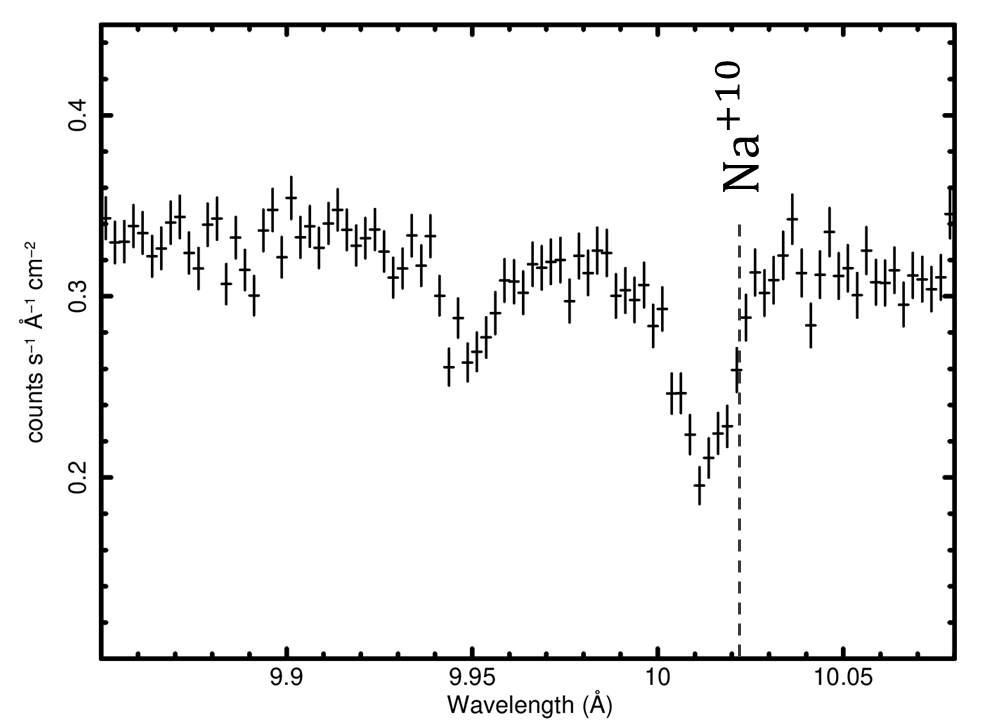}
    \includegraphics[width=0.32\textwidth]{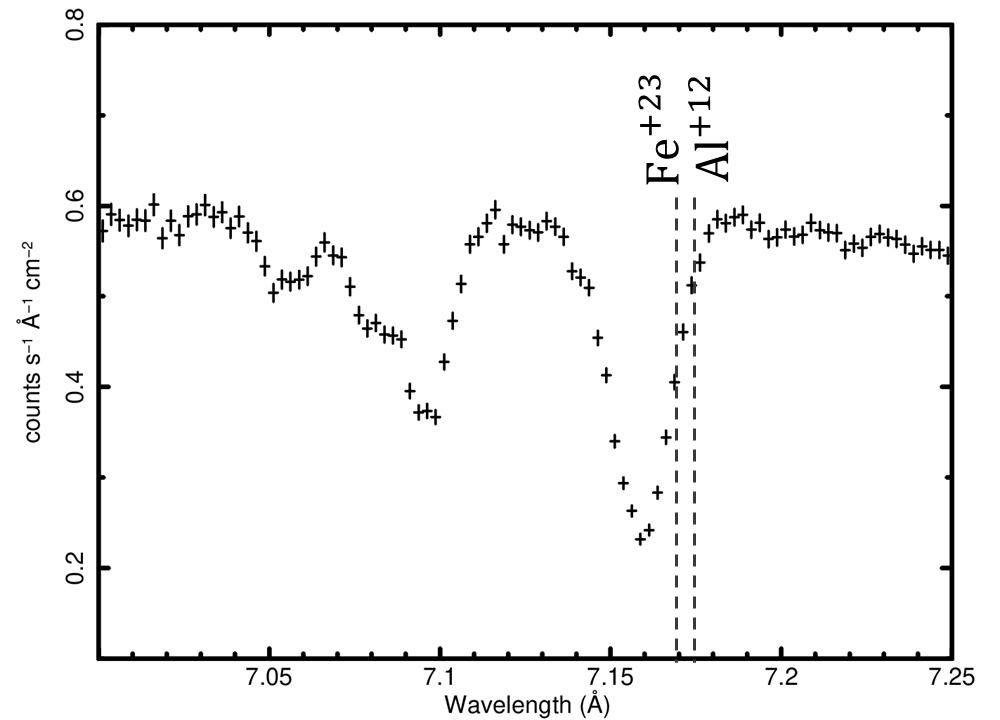}
    \includegraphics[width=0.32\textwidth]{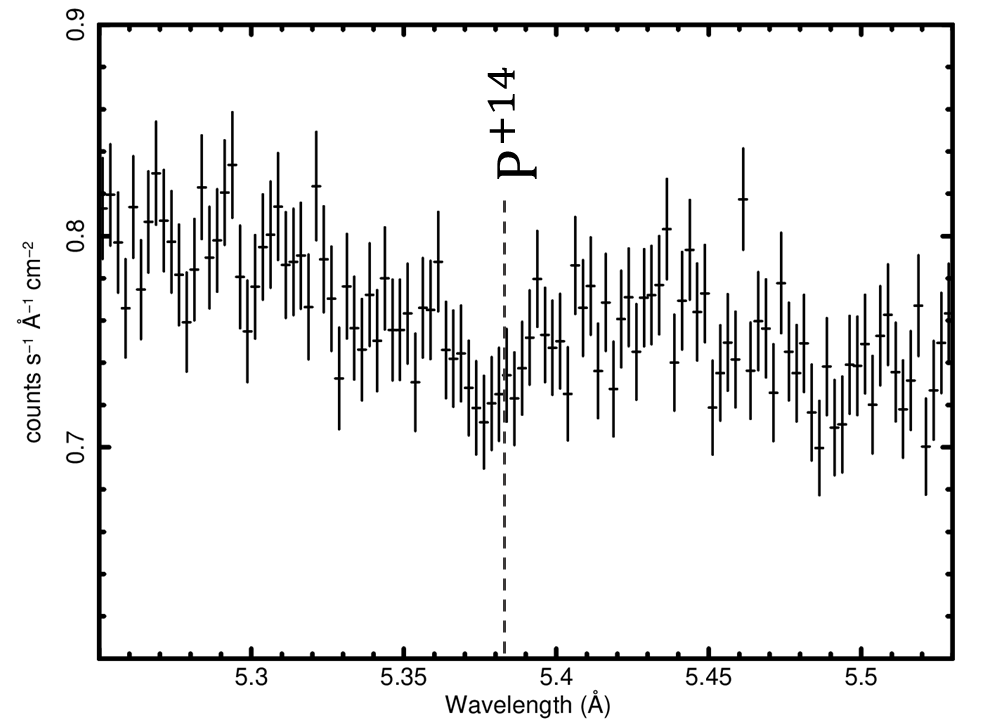}
    \includegraphics[width=0.32\textwidth]{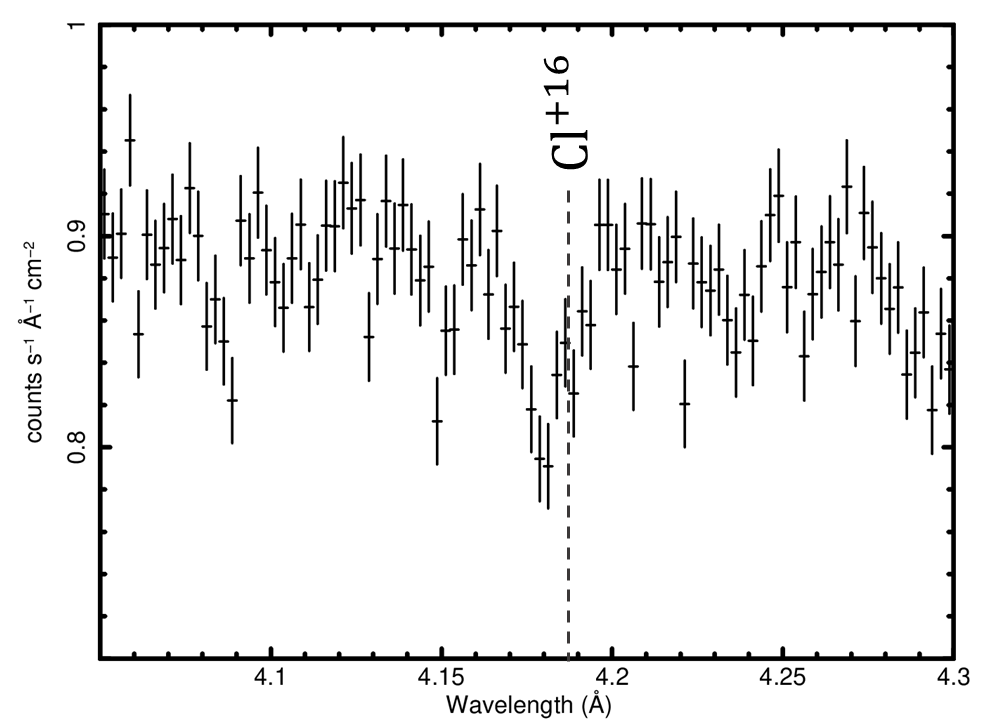}
    \includegraphics[width=0.32\textwidth]{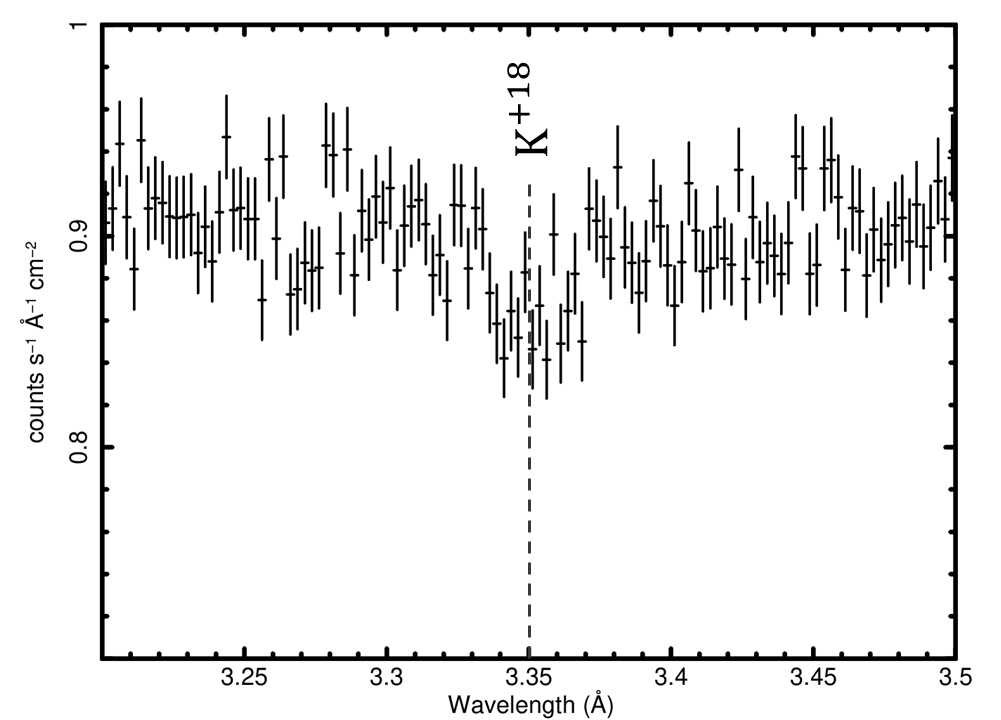}
    \caption{Segments from the \textit{Chandra}/HETG spectrum of the 2005 outflow of GRO\,J1655-40, zooming into the H-like Ly$\alpha$ lines of cosmically rare odd-Z elements. 
        Dashed lines mark laboratory wavelengths. }
    \label{fig:OddLines}
\end{figure}

\gro\ was observed with the \textit{Chandra}/HETG gratings during an outburst on 2005, April\,1. 
The observation, Id 5461, has a total exposure time of 26\,ks and a photon count of $4.2\times10^6$. The data were downloaded from the TGCAT archive \citep{Huenemoerder2011}
in fully reduced form.
The spectrum is uniquely rich between 1.5 - 20 \AA , and includes numerous absorption lines from 
eighteen different elements, many of which are rarely seen in astrophysical spectra, such as Na, Al, P, Cl, and K, see Fig.\,\ref{fig:OddLines}. 
The complex spectrum calls for the ion-by-ion and line-by-line analysis employed in this work. We are able to identify lines from H-like ions of sixteen different elements, He-like ions of eight elements and the L shell ions Fe$^{21}$-Fe$^{23}$, Ni$^{+25}$, and Co$^{+24}$. 


\section{Method} 
\label{sec:method}

Elemental abundances are measured through the relative strengths of the absorption lines, 
which are determined by the column density $N$, the amount of absorbing material expressed as an integral of density $n$ over the line of sight $N = \int ndr$. 
The ionic column densities $N_{\rm ion}$ are a product of the total (mostly H) column $N_{\rm H}$, the elemental abundance $A_{\rm Z}$, and the ion fractional abundance $f_{\rm ion}$.  
The outflowing gas includes a range of ionization ($\xi$), and each ion can emit from different $\xi$ regions. We therefore need to consider the contributions to $N_{\rm ion}$ over a range of $\xi$ values:

\begin{equation}
    \label{eq_Nion}
    N_{\rm ion} = A_{\rm Z} \int f_{\rm ion} \frac{dN_{\rm H}}{d\log \xi} d\log \xi
\end{equation}


\noindent where 
$dN_{\rm H} / d\log \xi$ is the Absorption Measure Distribution \citep[AMD,][]{Holczer2007}, and $f_{\rm ion}(\xi)$ values need to be taken from photoionization balance calculations, using the \gro\ ionizing spectrum \citep{Kallman2009}. $N_{\rm ion}$ is related to the line optical-depth profile through the relation $\tau(E) = \sigma (E) N_{\rm ion}$, where $E = h\nu$ is the line photon energy, and $\sigma (E)$ is the quantum mechanical line-absorption cross section.
The AMD function is recovered by requiring that it yields the correct $N_{\rm ion}$ in Eq.\,\ref{eq_Nion} for all ions.


\subsection{Ionic column densities}
\label{sec:Nion}

$N_{\rm ion}$ is measured using an ion-by-ion code 
that can fit simultaneously all lines of each ion,
and all ions at once \citep{Peretz2018}. 
The free parameters of the model are $N_{\rm ion}$ of each ion, and the global parameters of outflow velocity, and thermal and turbulent broadening. 

\begin{figure}
    \centering
    \includegraphics[width=\textwidth]{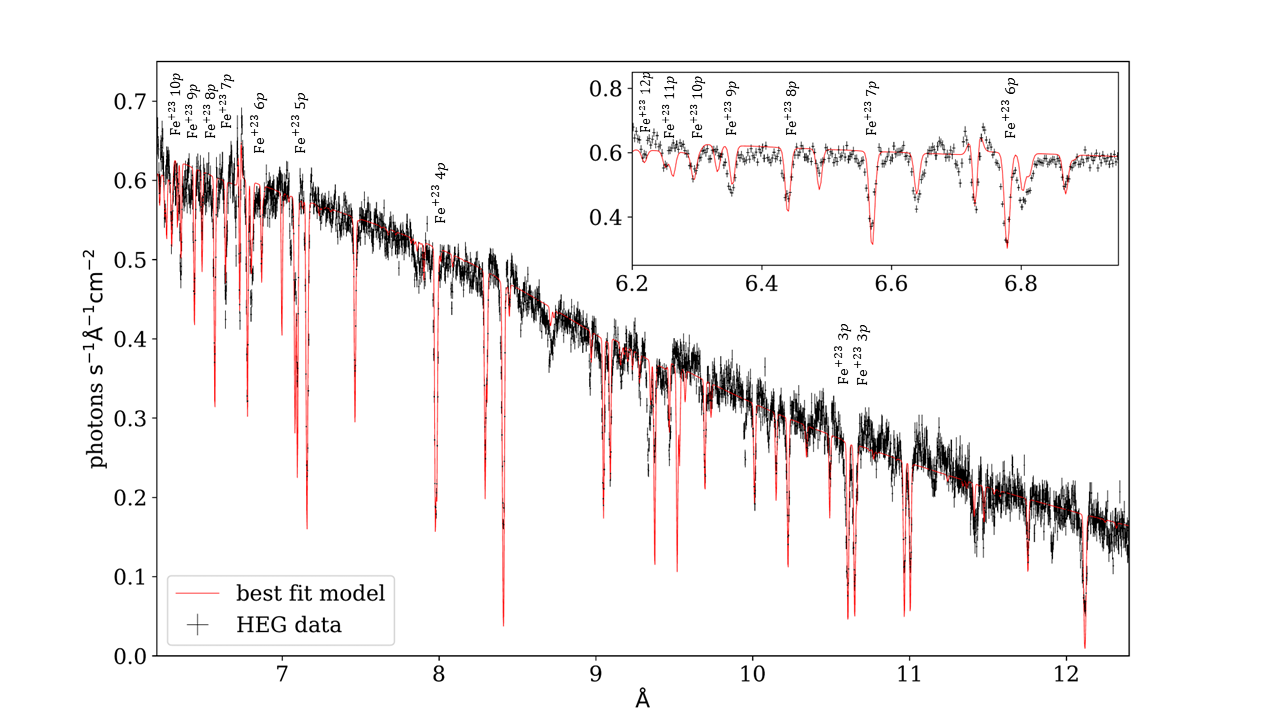}
    \caption{The 6.2-12.4\,\AA\ segment of the \hetg\ spectrum, fitted with an ion-by-ion absorption model. The series of Fe$^{+23}$ 2s-np lines is marked, and is fitted well by the model. N$_{\rm ion}$ values from this model are listed in Table\,\ref{tab:ions}.
    }
    \label{fig:ibifit}
\end{figure}

Fitting many lines from the same ion constrains the line broadening through all of their curves of growth,
even when they are unresolved. Additionally, fitting many ions together accounts for blends of lines of different ions.
Fig.\,\ref{fig:ibifit} shows our best fit ion-by-ion model 
in the 6.2-12.4 \AA\ band.
This is the richest part of the spectrum, which includes K-shell lines of Si, Al, Mg, Na, and Ne as well as L-shell lines of Fe, Ni and Co. The  (Balmer) series of Fe$^{+23}$ 2s-np lines is highlighted. Fitting so many observed lines of a single ion simultaneously tightly constrains $N_{\rm ion}$ (see Table\,\ref{tab:ions}). The atomic data are calculated with the HULLAC code \citep{Bar-Shalom2001} and corrected for laboratory wavelengths.


For Ly$\alpha$ lines of the rare odd-Z elements, which are not included in the \citet{Peretz2018} database, we measure individual-line equivalent widths. The equivalent width ($EW$) is the integral over the 
absorption line profile 
$EW = \int (1-e^{-\tau(E)})dE$, which is measured directly for each line. 
The relation between $EW$ and $N_{\rm ion}$ is called the curve of growth, and yields $N_{\rm ion}$. 
The advantage of individual line fitting is the superior constraint on the local continuum.

\begin{figure}
    \centering
    \includegraphics[width=\textwidth]{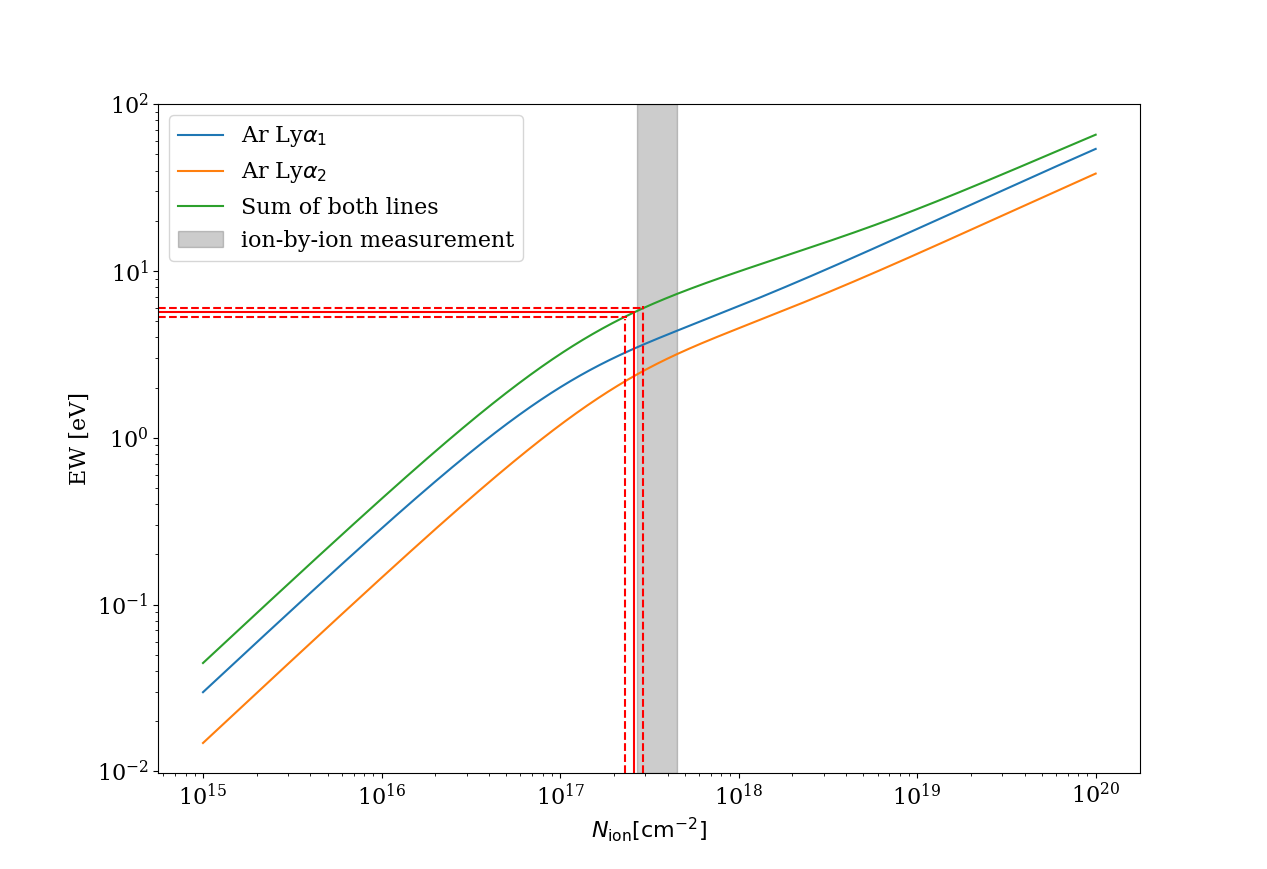}
    \caption{Theoretical curve-of-growth plot of the unresolved H-like Ar Ly$\alpha$ doublet. Each line of the Ly$\alpha$ doublet is calculated separately with its atomic data and for  $v_{\rm turb}$ = 70\,km\,s$^{-1}$, assuming a Voigt profile. The sum of both lines is also plotted, and then compared with the measured EW, represented by the horizontal solid red line; the dashed lines represent the margin of 90\% certainty, which then corresponds to the $N_{\rm ion}$ confidence region (vertical lines). $N_{\rm ion}$ found by the ion-by-ion fit (grey region) is consistent with the curve of growth analysis. 
    }
    \label{fig:EW}
\end{figure}

A theoretical curve of growth is calculated based on $\sigma(E)$ of each line which takes on the Voigt profile. The Voigt profile depends on the natural width and the measured Doppler velocity width of $v_{\rm turb} = \sqrt2 \sigma_g = 70\, \rm km\,s^{-1}$, where $\sigma_g$ is the Gaussian standard deviation.
$v_{\rm turb}$ is fitted with the ion-by-ion model. As an example, the curve of growth of Ar Ly$\alpha$, a deep and isolated doublet is given in Fig. \ref{fig:EW}. 
The flattening of the curve of growth is due to the saturation of the center of the line, where $\sigma(E)$ peaks, and absorption eliminates 100\% of the flux. This flattening results in relatively large uncertainties in $N_{\rm ion}$ (and consequently the elemental abundances), even when the $EW$ is accurately measured (see Fig. \ref{fig:EW}). 
All currently measured lines are listed in Table\,\ref{tab:lines}, along with their identified ions and $EW$.

Wherever possible, $N_{\rm ion}$ was measured with ion-by-ion fitting (Fig.\,\ref{fig:ibifit}) and directly (Fig.\,\ref{fig:EW}) 
to validate consistency. The values quoted in Table\,\ref{tab:ions} are those measured with ion-by-ion fitting, when available.
Only for Mg$^{+11}$ and Si$^{+13}$, the ion-by-ion fit underestimates the direct measurement, and the $N_{\rm ion}$ values quoted (Table\,\ref{tab:ions}) are those measured directly. These discrepancies are mainly due to the semi-global fit not accurately capturing the local continuum. Although fitting a series of lines together can provide stronger constraints on N$_{\rm ion}$, the local continuum may deviate from the data, introducing uncertainties. Fitting line profiles individually allows for a better fit to the continuum.

\begin{figure}
    \centering
    \includegraphics[width=\textwidth]{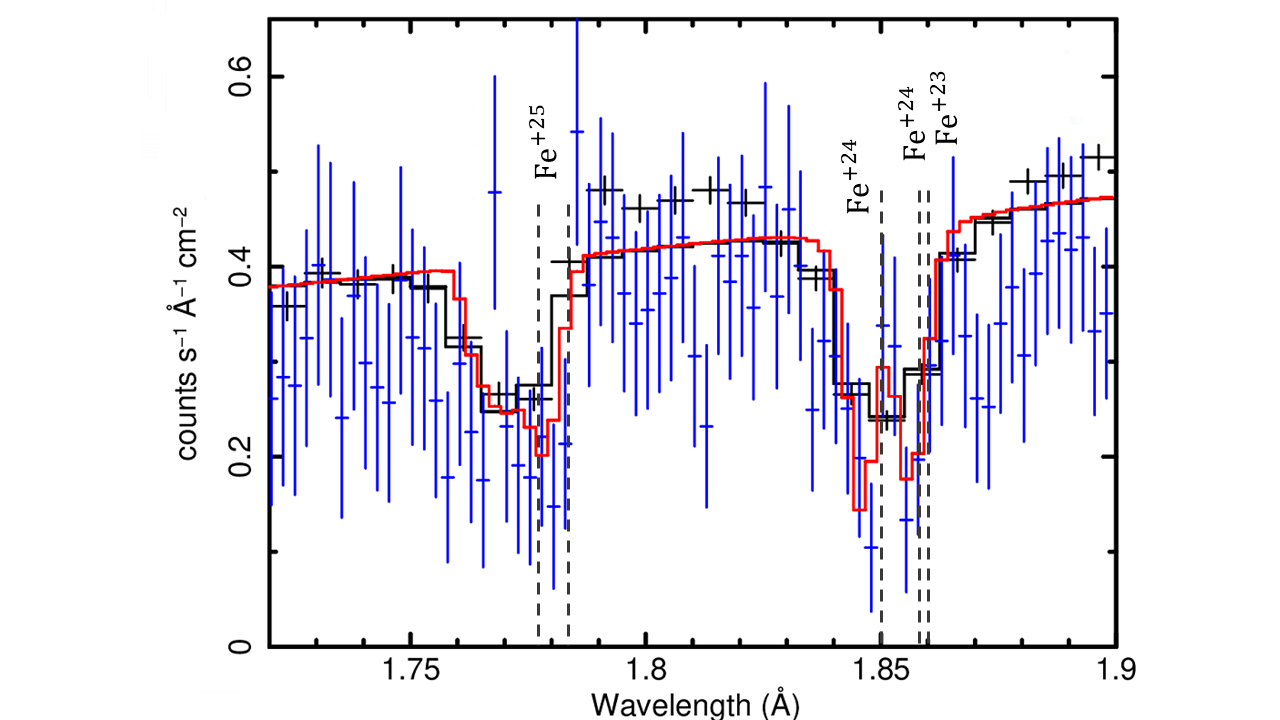}
    \caption{HEG first- (black) and third- (blue data, model in red) order spectra. 
    The rest wavelengths (left to right, respectively) of Fe$^{+25}$ Ly$\alpha$ doublet, Fe$^{+24}$ He$\alpha$ $r$ and $i$, and Fe$^{+23}$ K$\alpha$ are marked. The latter two are blended even in third order.  Note there are differences in the continuum between orders, which is likely an issue of the absolute calibration at these extreme low wavelengths.}
    \label{fig:Fe3rd}
\end{figure}

All lines in the spectrum share approximately the same kinematics of blue-shift of $-400\, \rm km\,s^{-1}$ and broadening of $v_{\rm turb}=70\, \rm km\,s^{-1}$ expect H-like Fe). 
The third-order HETG spectrum resolves the Fe$^{+24}$ resonance line ($r$) from the intercombination ($i$) line ($r$ at 1.85\AA\ and $i$ at 1.86\AA ), which is blended with the Fe$^{+23}$ K$\alpha$ line, see Fig.\,\ref{fig:Fe3rd}.    
When fitting these lines, we allowed separate continua for the two orders, but confirmed that all lines $EW$s are consistent. This allows for an accurate measurement of the Fe$^{+24}$ ion column density. $N_{\rm ion}$ for $\rm Fe^{+23}$ is measured by fitting the series of its 2s-np lines (see Fig.\,\ref{fig:ibifit}). Since Fe$^{+25}$ has a much higher outflow velocity, at $\sim -1200\, \rm km\,s^{-1}$, and $v_{\rm turb}=1300\, \rm km\,s^{-1}$, it is likely a separate component of the outflow \citep{Kallman2009}, and we exclude it from the AMD analysis. Fitting the Fe$^{+25}$ doublet with two Voigt profiles in the third order confirms these higher outflow and turbulent velocities. All final $N_{\rm ion}$ values are detailed at Tabel\,\ref{tab:ions}, and plotted in the top panel of Fig.\,\ref{fig:FullAMD}.

\subsection{Reconstructing the AMD}
\label{sec:amd}

\begin{figure}
    \centering
    \includegraphics[width=\textwidth]{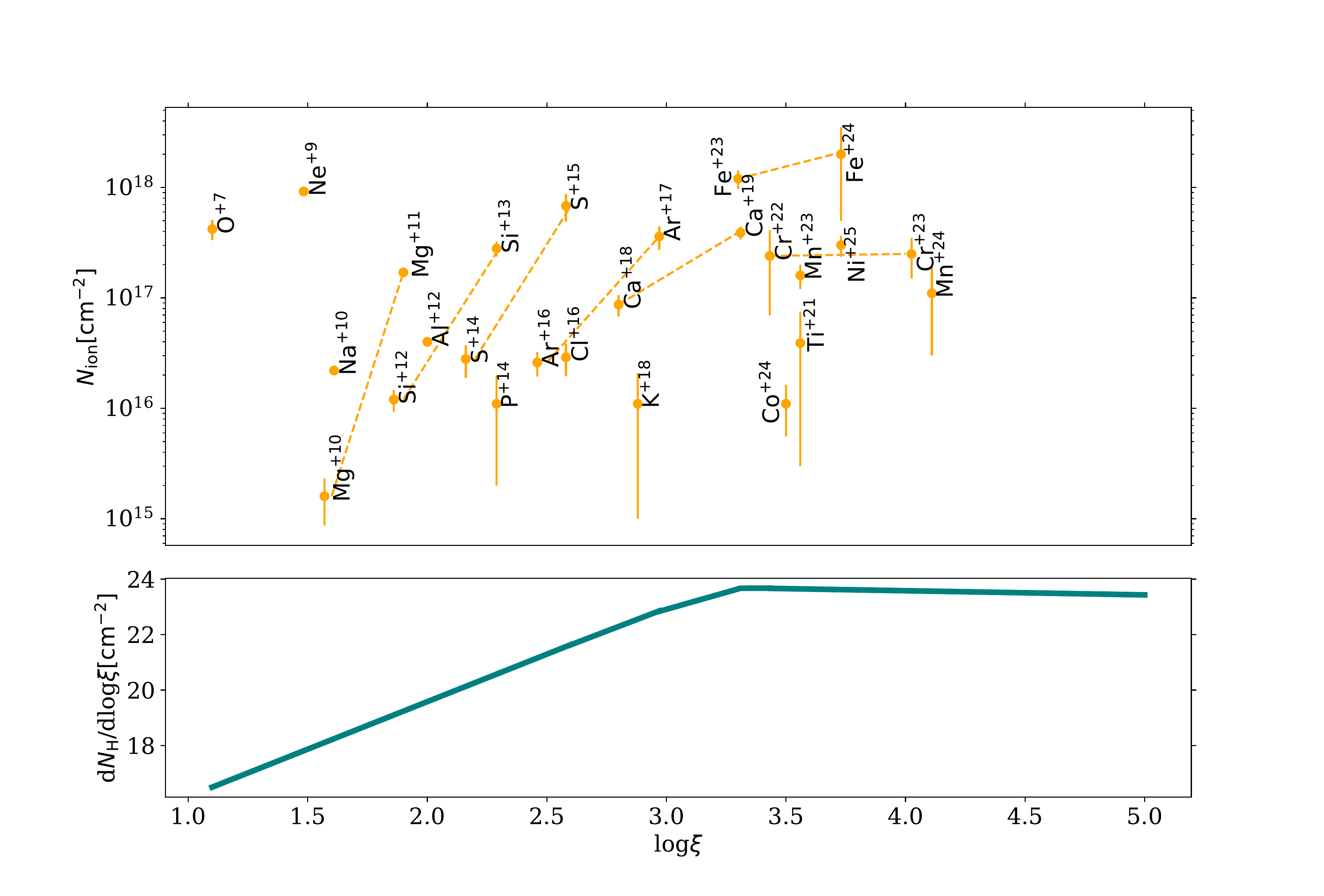}
    \caption{(\textit{top panel}) Distribution of ionic column densities $N_{\rm ion}$, each plotted at its $\xi$ of maximal formation $\xi_{\rm max}$, see Table \ref{tab:ions}. 
    This outlines the overall distribution of $N_{\rm H}$ with $\xi$ in the \gro\ outflow. Dashed lines represent the local slopes from H-like and He-like ion pairs, which are then used to produce the continuous distribution, i.e. the AMD, (\textit{bottom panel}) 
    With no actual H in the spectrum, H-like Ca is arbitrarily chosen as the reference point. 
    The AMD drastically rises up to $\log\xi = 3.3$, and slightly decreases towards high $\xi$. The integral over the AMD yields total column density of $6.2\times10^{23}$ cm$^{-2}$.
    }
    \label{fig:FullAMD}
\end{figure}

Obtaining relative elemental abundances from the various $N_{\rm ion}$ values requires considering that $N_{\rm H}$ in Eq.\,(\ref{eq_Nion}) can depend on the ionization parameter $\xi$. The functional form of the AMD is unknown apriori. Here,
the AMD is constructed from local 
segments of 
power-laws by requiring that they give the same element abundance $A_{\rm Z}$ for all pairs of H-like and He-like ions (see Eq.\,\ref{eq_Nion}). The individual segments below $\log \xi = 3.3$ all indicate a steeply rising trend. 
Hence for simplicity, we assumed a broken power-law AMD (Fig.\,\ref{fig:FullAMD}) between $\log \xi = 0 - 5$, where the slopes and break points are the free parameters.
By iterations, each time obtaining $A_Z$ from Eq.\,\ref{eq_Nion} for all ions, we seek the best ($A_Z$) match between ion pairs.
We use $N_{\rm ion}$ measured values from Sec. \ref{sec:Nion}, and $f_{\rm ion}(\xi)$ from \citet{Kallman2009} in Eq.\,\ref{eq_Nion}.
This process eventually results in the AMD presented in the bottom panel of Fig.\,\ref{fig:FullAMD} and the best abundance estimates.

In Fig.\,\ref{fig:FullAMD} (top panel), we plot measured $N_{\rm ion}$ values for all ions in the spectrum, where each ion is plotted at its maximal formation position (log$\xi_{\rm max}$, see Table\,\ref{tab:ions}).
The relative positions of ion pairs on the (top) plot of Fig.\,\ref{fig:FullAMD} indicate the local slope of the AMD 
hinting at a steep distribution. 
The bottom panel of Fig.\,\ref{fig:FullAMD} shows the AMD that best produces the $N_{\rm ions}$ values in the top panel. It sharply increases with $\xi$ up to $\log \xi = 3.3$ and then levels off or even slightly decreases.
There are no ion pairs below $\log\xi = 1.6$ or above $\log\xi = 4.0$, hence the AMD is less tightly constrained in these regions. 
The fact that we see no sign of decreasing AMD towards high ionization is consistent with previous models \citep{Shidatsu2016, Neilsen2016, Fukumura2017} that require fully ionized gas in the inner parts of the wind, which does not absorb in lines, but makes it Compton-thick. In the case that the AMD extends up to $\log\xi = 5$ we get a total column density of $N_{\rm H} = 6.2\times10^{23}$ cm$^{-2}$ that does not include the fast H-like Fe component. This is consistent with the models of \citet{Kallman2009}. \citet{Tomaru2023} report $N_{\rm H}$ that is an order of magnitude higher but it refers to the fast H-like Fe component. They also invoke partial covering of the X-ray source, which exponentially raises $N_{\rm H}$.

The 
AMD of the wind 
points to two different regions. 
A high-ionization flat component at $\log \xi > 3.3$ and a low-ionization steep one (Fig.\,\ref{fig:FullAMD}).
High ionization is typical of X-ray binary outflows, while low ionization is rare \citep{Trigo2016}. 
It has been shown that the slope of the AMD 
$\sim \xi ^a$ maps exactly the line-of-sight radial wind-density profile $n(r)$ \citep[][Eq.\,7 therein]{Behar2009}. 
The high-ionization flat ($a \approx 0$) distribution of \gro\ is akin to those observed in active galactic outflows \citep{Keshet2022}.
Such an AMD is produced by a shallow $n(r)\propto r^{-1}$ density profile, which can be due to a magnetic disk wind 
\citep{Fukumura2017}. 
Further out, the wind interacts with circumstellar gas, and becomes more spherical, approaching $n(r)\propto r^{-2}$ \citep{Fukumura2010}.
The steep AMD ($\log \xi < 3$) could be this more extended region, where $n$ decreases more strongly with $r$.

\section{Elemental Abundances} \label{sec:results}

Once the AMD profile is solved in terms of $N_{\rm H}$, the relative abundances are determined by the factor $A_Z$ applied to N$_{\rm ion}$ that is required to match the AMD (see Eq.\,\ref{eq_Nion}). For each element, the abundance measurement is taken as the weighted average from all available ions. 
This vertical scaling between the various elements 
represents their relative abundances. The final abundances of eighteen elements in the outflow of \gro\ are listed in Table\,\ref{tab:abund}. 
As far as we know, this sets a record for any astrophysical measurement, mostly because the visible and UV bands do not contain spectral lines of as many elements as the soft X-ray band. Abundances in Table\,\ref{tab:abund} are given relative to their solar values \citep{Asplund2009}.
The X-ray measured abundances of the wind are compared in the table to the photospheric values of the companion \citep{Gonzales2008}, and to the values previously measured with a global fit \citep{Kallman2009}.
Since no H lines can be observed in X-rays, and 
since the Ca/H abundance ratio is solar in the photosphere, all abundances in Table\,\ref{tab:abund} are listed assuming Ca is solar. 

\begin{deluxetable}{lccccccc}
\tabletypesize{\scriptsize}
\tablewidth{0pt}
\tablecaption{Ions identified in the HETG spectrum of \gro\ and used for abundance estimates. Values in the last two columns are taken from photo-ionization models \citep{Kallman2009}. All uncertainties represent 90\% confidence. 
}
\label{tab:ions}
\tablehead{
\colhead{Ion} 
& \colhead{$N_{\rm ion}$} 
& \colhead{log$\xi_{\rm max}$ } 
& \colhead{$ f_{\rm ion}(\xi_{\rm max})$ }\\
\colhead{} 
& \colhead{$10^{17}$[cm$^{-2}$]} 
& \colhead{[erg $\,\rm s^{-1}$cm]} 
& \colhead{} 
}
\startdata
{$\rm O^{+7}$}   & $1.1 \pm 1.0$    & 1.1 & 0.54 \\
{$\rm Ne^{+9}$}  & $4.2 \pm 4.0$    & 1.5 & 0.55 \\
{$\rm Na^{+10}$} &  $0.22 \pm 0.09$ & 1.6 & 0.66 \\             
{$\rm Mg^{+11}$} & $2.3 \pm 0.2$    & 1.9 & 0.53 \\
{$\rm Mg^{+10}$} & $0.016 \pm 0.007$& 1.6 & 0.27 \\
{$\rm Al^{+12}$} & $0.40 \pm 0.04$  & 2.0 & 0.64 \\
{$\rm Si^{+13}$} & $8 \pm 2$        & 2.3 & 0.55 \\
{$\rm Si^{+12}$} & $0.3 \pm 0.1$    & 1.9 & 0.30 \\
{$\rm P^{+14}$}  & $0.11 \pm 0.10$  & 2.3 & 0.66 \\
{$\rm S^{+15}$}  & $11 \pm 3$       & 2.6 & 0.54\\
{$\rm S^{+14}$}  & $0.22 \pm 0.05$  & 2.2 & 0.34 \\
{$\rm Cl^{+16}$} & $0.3 \pm 0.1$    & 2.6 & 0.69 \\
{$\rm Ar{+17}$}  & $2.6 \pm 0.3$    & 3.0 & 0.52 \\
{$\rm Ar^{+16}$} & $0.24 \pm 0.04$  & 2.5 & 0.41 \\
{$\rm K^{+18}$}  & $0.11 \pm 0.10$  & 2.9 & 0.71 \\
{$\rm Ca^{+19}$} & $3.6 \pm 0.9$    & 3.3 & 0.55 \\
{$\rm Ca{+18}$}  & $0.5 \pm 0.1$    & 2.8 & 0.41 \\
{$\rm Ti{+21}$}  & $0.39 \pm 0.36$  & 3.5 & 0.67 \\
{$\rm Cr{+23}$}  & $3.0 \pm 1.0$    & 4.0 & 0.67\\
{$\rm Cr^{+22}$} & $2.4 \pm 1.7$    & 3.4 & 0.35 \\
{$\rm Mn^{+24}$} & $1.1 \pm 0.5$    & 4.1 & 0.71 \\
{$\rm Mn^{+23}$} & $1.6 \pm 0.4$    & 3.6 & 0.37 \\
{$\rm Fe^{+25}$} & $9.0 \pm 1.0$    & 4.3 & 0.52 \\
{$\rm Fe^{+24}$} & $20 \pm 15$      & 3.7 & 0.69 \\
{$\rm Fe^{+23}$} & $12 \pm 2$       & 3.3 & 0.19 \\
{$\rm Co^{+24}$} & $0.11 \pm 0.06$  & 3.6 & 0.14 \\
{$\rm Ni^{+25}$} & $3.0 \pm 0.6$    & 3.7 & 0.18\\
\enddata

\end{deluxetable}

The abundances obtained by a different analysis approach of the same X-ray spectrum were published by \citet{Kallman2009}. Most of our abundances agree with their results within the uncertainties. The significant discrepancies up to a factor of 2.8 are for low-Z (atomic number) elements below P (Z$<15$). This is a result of the systematic difference of the two approaches. \citet{Kallman2009} used a single ionization component with $\log\xi = 4$, where the fractional abundances of low-Z ions are. Therefore, high elemental abundances were required to fit the spectrum. The current analysis with an AMD extending to the low $\xi$ region fits the spectrum with lower abundances. 
The ion-by-ion method allows more flexibility in fitting the spectrum, particularly weak lines that the global fit is not sensitive to.
This results for the most part in smaller uncertainties and well constrained odd-Z abundances,
 which are important in comparing with SN models (see Sec.\, \ref{sec:discussion}). 
\citet{Kallman2009} mentioned a possible thermal instability at $\log\xi \approx  2$. Considering the same AMD reconstructed in Sec.\,\ref{sec:amd} with a cut-off at $\log \xi = 2$ produces the same measured abundances, with negligible differences.
 
Abundance measurements from the companion atmosphere based on optical observations \citep{Gonzales2008} are also listed in Table\,\ref{tab:abund}. 
The abundances of O, Mg, Si, and S, which are supersolar in the photosphere, are subsolar in the wind. An earlier photospheric measurement reports Mg, Si, and S solar abundances \citep{Foellmi2007}.

\begin{deluxetable}{lccccccc}
\tabletypesize{\scriptsize}
\tablewidth{0pt}
\tablecaption{Lines measured in the HETG spectrum of \gro\ and used for abundance estimates. All uncertainties are 90\% confidence. Laboratory energies are taken from the NIST database \citep{NIST_ASD}.
}
\label{tab:lines}
\tablehead{
\colhead{Line} & \colhead{Measured Energy}
& \colhead{Laboratory Energy}
& \colhead{$EW$} \\
\colhead{} & \colhead{[keV]} & \colhead{[keV]}  & \colhead{$10^{-3}$[keV]}
}
\startdata
{$\rm O^{+7}$ Ly$\alpha$}   & 0.6547 & 0.6537 & $0.8 \pm 0.3$ \\
{$\rm O^{+7}$ Ly$\beta$}    & 0.7756 & 0.7747 & $0.7 \pm 0.1$ \\
{$\rm Ne^{+9}$ Ly$\alpha$}  & 1.0238 & 1.0218 & $2 \pm 1$     \\
{$\rm Ne^{+9}$ Ly$\beta$}   & 1.2124 & 1.2111 & $1.38 \pm 0.05$ \\
{$\rm Na^{+10}$ Ly$\alpha$} & 1.2382 & 1.2368 & $0.7 \pm 0.2$ \\
{$\rm Na^{+10}$ Ly$\beta$}  & 1.4680 & 1.4658 & $0.21 \pm 0.04$ \\               
{$\rm Mg^{+11}$ Ly$\alpha$} & 1.4743 & 1.4723 & $2.45 \pm 0.06$ \\
{$\rm Mg^{+11}$ Ly$\beta$}  & 1.7480 & 1.7450 & $2.0 \pm 0.8$ \\
{$\rm Mg^{+10}$ He$\alpha$} & 1.3533 & 1.3538 & $0.13 \pm 0.12$ \\
{$\rm Al^{+12}$ Ly$\alpha$} & 1.7318$^*$ & 1.7285 & $2.69 \pm 0.04$ \\
{$\rm Si^{+13}$ Ly$\alpha$} & 2.0087 & 2.0054 & $5.2 \pm 0.2$ \\
{$\rm Si^{+13}$ Ly$\beta$}  & 2.3809 & 2.3768 & $2.6 \pm 0.2$ \\
{$\rm Si^{+12}$ He$\alpha$} & 1.8673 & 1.8671 & $1.3 \pm 0.5$ \\
{$\rm P^{+14}$ Ly$\alpha$}  & 2.3061 & 2.3031 & $0.46 \pm 0.45$  \\
{$\rm P^{+14}$ Ly$\beta$}   & 2.7340 & 2.7296 & $0.11 \pm 0.10$  \\
{$\rm S^{+15}$ Ly$\alpha$}  & 2.6249 & 2.6215 & $8.0 \pm 0.7$    \\
{$\rm S^{+15}$ Ly$\beta$}   & 3.1122 & 3.1069 & $2.8 \pm 0.3$     \\
{$\rm S^{+14}$ He$\alpha$}  & 2.4649 & 2.4635 & $1.3 \pm 0.2$    \\
{$\rm Cl^{+16}$ Ly$\alpha$} & 2.9659 & 2.9608 & $1.2 \pm 0.3$    \\
{$\rm Cl^{+16}$ Ly$\beta$}  & 3.5050 & 3.5090 & $0.8 \pm 0.5$  \\
{$\rm Ar{+17}$ Ly$\alpha$}  & 3.3262 & 3.3209 & $5.7 \pm 0.3$    \\
{$\rm Ar{+17}$ Ly$\beta$}   & 3.9404 & 3.9359 & $2.6 \pm 0.5$   \\
{$\rm Ar^{+16}$ He$\alpha$} & 3.1447 & 3.1435 & $1.5 \pm 0.2$    \\
{$\rm K^{+18}$ Ly$\alpha$}  & 3.700  & 3.7020 & $0.5 \pm 0.4$    \\
{$\rm Ca^{+19}$ Ly$\alpha$} & 4.1123 & 4.1042 & $8 \pm 1$      \\
{$\rm Ca{+19}$ Ly$\beta$}   & 4.8715 & 4.8642 & $4.1 \pm 0.8$      \\
{$\rm Ca{+18}$ He$\alpha$}  & 3.9086 & 3.9076 & $3.1 \pm 0.4$     \\
{$\rm Ti{+21}$ Ly$\alpha$}  & 4.9739 & 4.9718 & $1.6 \pm 1.4$     \\
{$\rm Cr{+23}$ Ly$\alpha$}  & 5.9381 & 5.9243 & $8 \pm 2$     \\
{$\rm Cr^{+22}$ He$\alpha$} & 5.6902 & 5.6907 & $10 \pm 4$    \\
{$\rm Mn^{+24}$ Ly$\alpha$} & 6.4440 & 6.4326 & $4 \pm 2$    \\
{$\rm Mn^{+24}$ Ly$\beta$}  & 7.6120 & 7.6243 & $8 \pm 5$  \\
{$\rm Mn^{+23}$ He$\alpha$} & 6.1770 & 6.1901 & $8 \pm 1$  \\
{$\rm Fe^{+25}$ Ly$\alpha$} & 7.0047$^{**}$ & 6.9693 & $30 \pm 10$     \\
{$\rm Fe^{+24}$ He$\alpha$} & 6.6990 & 6.7112 & $50 \pm 20$    \\
{$\rm Co^{+24}$ K$\alpha$}  & 1.2673 & 1.2660 & $0.3 \pm 0.1$    \\
{$\rm Ni^{+25}$ K$\alpha$1}& 1.3702  & 1.3685 & $1.9 \pm 0.1$       \\
{$\rm Ni^{+25}$ K$\alpha$2}& 1.3636  & 1.3620 & $1.0 \pm 0.3$       \\
\enddata
\tablenotetext{*}{
 blended lines}
\tablenotetext{**}{
 higher velocity component}

\end{deluxetable}



\begin{deluxetable}{lccc}
\tabletypesize{\scriptsize}
\tablewidth{0pt}
\tablecaption{Elemental abundances in the X-ray wind of \gro\ compared to those measured in the visible waveband for the companion star \citep{Gonzales2008} and to those measured in a global fit \citep{Kallman2009}.
X-ray measurements only provide relative abundances, and are thus  listed relative to Ca, while companion measurements are relative to H, but normalized here relative to Ca for the comparison. 
All abundances are given with respect to their solar values.
}
\label{tab:abund}
\tablehead{
\colhead{Element} & \colhead{Measured $A_Z$/Ca} 
& \colhead{Companion $A_Z$/Ca} & \colhead{Global fit $A_Z$/Ca} 
}
\startdata
{O}  &  $0.5\pm0.1$   &  $9\pm4$      & $1.4\pm0.9$   \\ 
{Ne} &  $1.9\pm0.3$   &  -            & $5.4\pm0.9$   \\    
{Na} &  $1.0\pm0.1$   &  $2.5\pm1.8$  & $0.9\pm0.2$   \\
{Mg} &  $0.27\pm0.04$ &  $2.9\pm1.4$  & $0.55\pm0.05$ \\
{Al} &  $0.48\pm0.08$ &  $1.2\pm0.6$  & $0.8\pm0.2$   \\
{Si} &  $0.23\pm0.04$ &  $4\pm2$      & $0.6\pm0.1$   \\
{P}  &  $0.7\pm0.6$   &  -            & $0.0\pm0.2$   \\
{S}  &  $0.6\pm0.1$   &  $6\pm2$      & $0.8\pm0.2$   \\
{Cl} &  $0.8\pm0.3$   &  -            & $0.10\pm0.09$ \\
{Ar} &  $0.7\pm0.2$   &  -            & $1.0\pm0.2$   \\
{K}  &  $0.57\pm0.53$ &  -            & $0.50\pm0.45$ \\
{Ca} &  $1.0\pm0.2$   &  $1.0\pm0.5$  & $1.0\pm0.9$   \\
{Ti} &  $1.6\pm1.5$   &  $2.0\pm1.0$  & $1.1\pm0.5$   \\
{Cr} &  $2.2\pm0.9$   &  -            & $2.6\pm2.5$   \\
{Mn} &  $2.6\pm0.8$   &  -            & $14\pm12$     \\
{Fe} &  $0.6\pm0.3$   &  $0.8\pm0.3$  & $0.96\pm0.05$ \\
{Co} &  $3\pm2$       &  -            & $7\pm6$       \\
{Ni} &  $4\pm1$       &  $1.1\pm0.6$  & $0.61\pm0.05$ \\
\enddata
\end{deluxetable}

\section{Discussion} \label{sec:discussion}

\begin{figure}
    \centering
    \includegraphics[width=\textwidth]{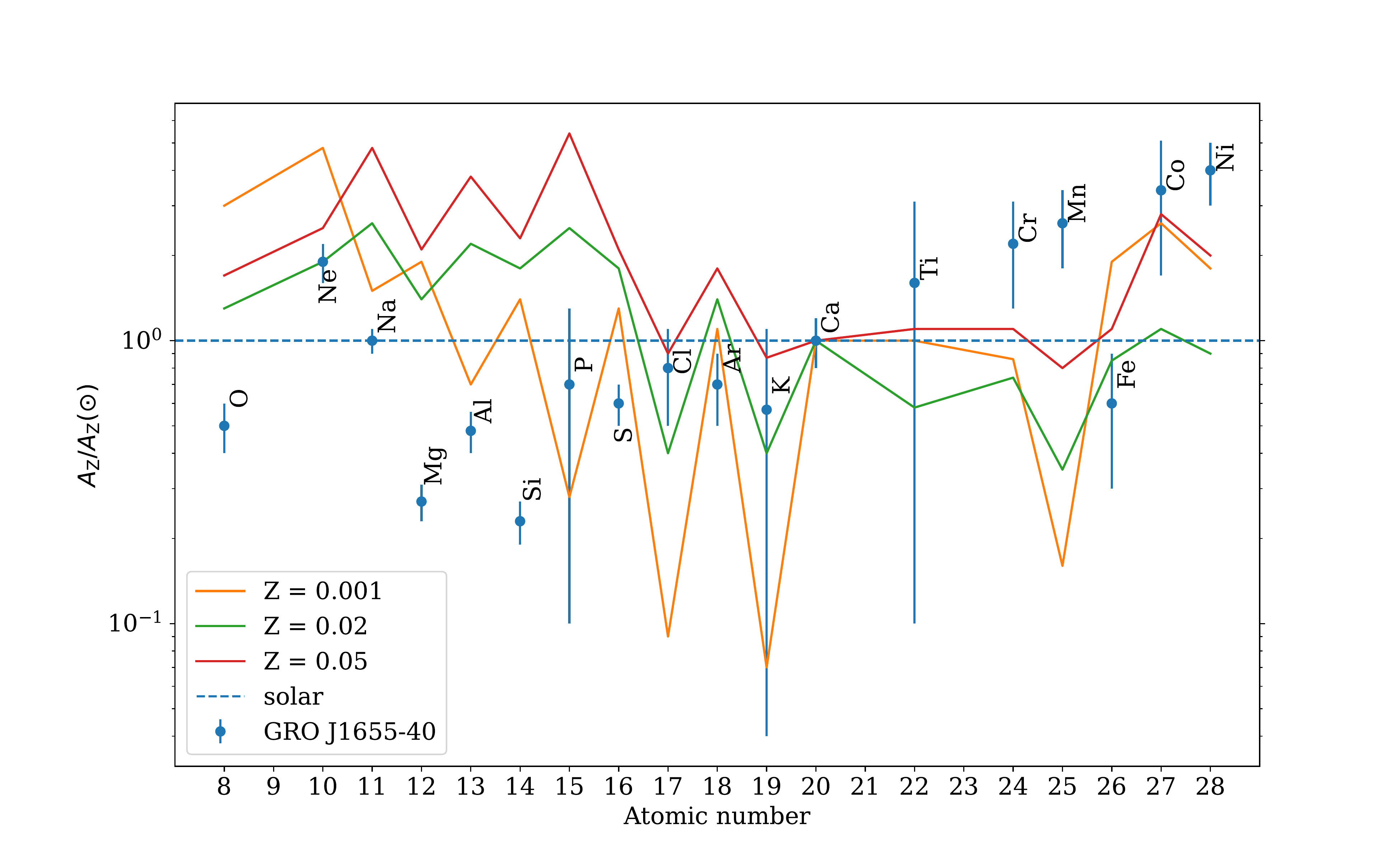}
    \caption{ Comparison of 
    the presently measured abundances in the \gro\ outflow to different SN model yields 
    from  \citet{Nomoto2013}, for a $25\,M_\odot$ progenitor with different metallicity values, given in fractional mass units; Z=0.02 is solar metallicity.
    The high relative abundances of Na, Al, P, Cl, Mn, and Co indicate a tendency towards models with a progenitor of solar metallicity or higher. 
    Note that all data points are scaled to a solar Ca abundance, but can be (uniformly) vertically shifted.
    }
    \label{fig:SNmodelsZ}
\end{figure}

\begin{figure}
    \centering
    \includegraphics[width=\textwidth]{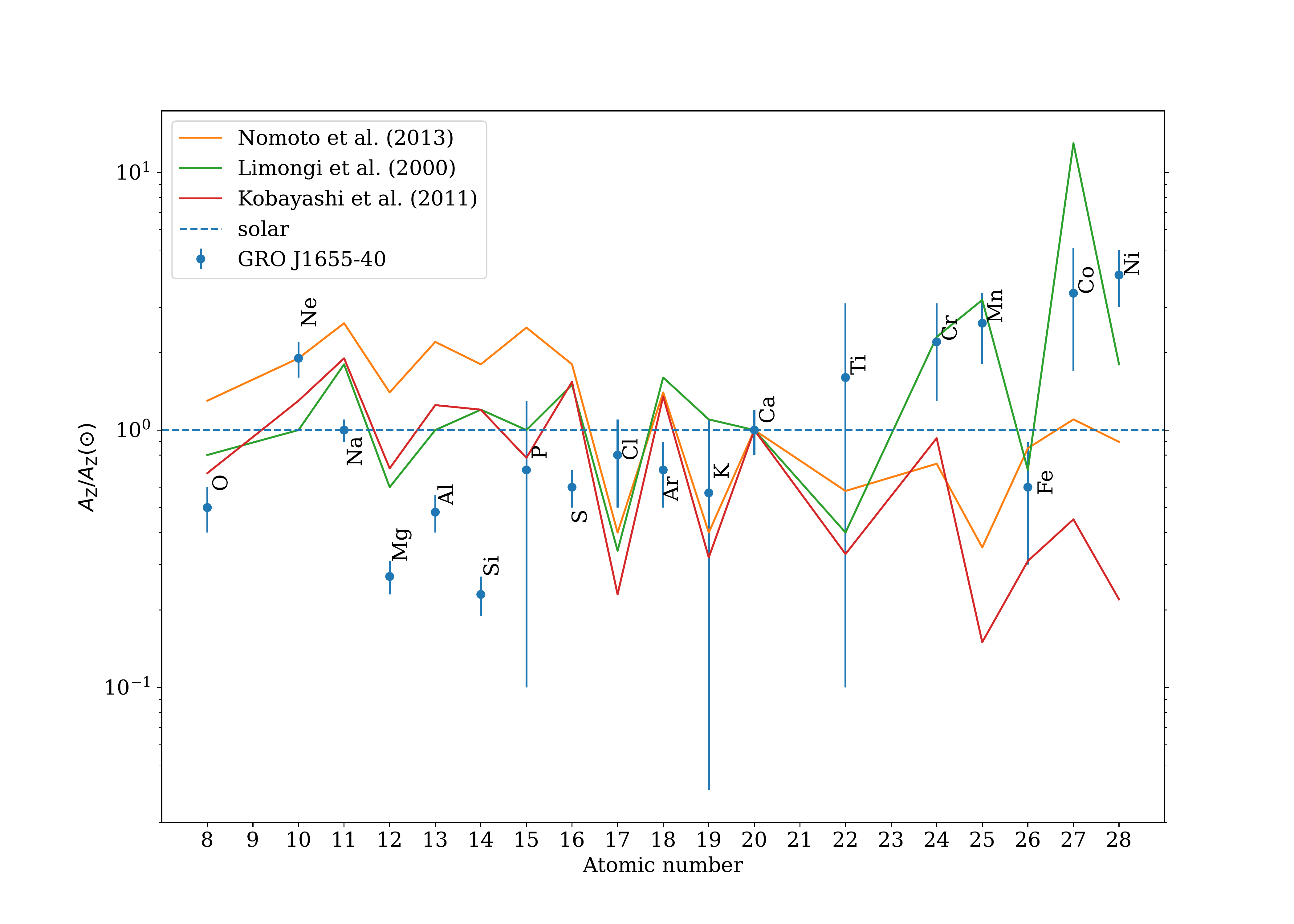}
    \caption{ Three different SN model yields for a $25\,M_\odot$ progenitor with solar metallicity ($Z = 0.02$), demonstrating the theoretical uncertainties of such models, none of which is in perfect agreement with the \gro\ abundances. The small peaks of odd-Z elements predicted by \citet{Nomoto2013} qualitatively follow the measurements. Statistically, the yields of \citet{Limongi2000} give a somewhat better agreement. }
    \label{fig:SNmodels}
\end{figure}

It is puzzling that the \gro\ wind elemental abundances are significantly different from those of the companion star. This could be a result of fractionation in the accretion process or contamination by ambient material. We compare the measured abundances with SN model yields of massive stars ($\geq25 M_\odot$), the progenitor candidates of the BH. Fig.\,\ref{fig:SNmodelsZ} presents a comparison of the measured abundances with core-collapse SN model yields from \citet{Nomoto2013}, for 25\,$M_\odot$ stars of different metallicites. Models with higher mass progenitors are less compatible with our results, therefore herewith we focus on $25\,M_{\odot}$ models.
The measured elemental abundances of \gro , and in particular the showing of the odd-Z elements hint to the wind being contaminated by debris from a SN explosion of a $25\,M_{\odot}$ star, with solar or higher metallicity.
Odd-Z elements are produced 
from even-Z proton-rich isotopes by electron capture, or positron decay \citep{Nomoto2013}. The present peaks of odd-Z abundances Na, Al, P, Cl, Mn, and Co (Fig.\,\ref{fig:SNmodelsZ}) confirm the role of neutronization processes, which is an important component in understanding the abundances of elements in the universe.
Different SN models vary in their abundance yields. Fig.\,\ref{fig:SNmodels} shows three models for the same progenitor mass and metallicity. 
Since the \gro\ abundance pattern does not exactly match the SN models, we suspect the wind is a mixture of the SN debris and the companion composition. 

High mass SNe are predicted to leave behind a BH of only $\sim3\,M_\odot$, with $\sim 2\,M_\odot$ falling back \citep{Israelian1999, Podsiadlowski2002, Gonzales2008}, while ejecting most of their mass into the circumstellar medium \citep{Nomoto2013}.
Indeed, the total mass of the \gro\ binary is $<10\,M_\odot$, leaving at least $15\,M_\odot$ in the ambient circumstellar region.
Corroborating evidence for this excess mass comes from the optical/infrared thermal emission that coincided with the \gro\ outburst \citep{Neilsen2016, Shidatsu2016}. This excess was ascribed by \citet{Neilsen2016} to a cool, Compton thick photosphere, 
with a size of $\sim 5\times10^{11}$\,cm, \citep[c.f.,][]{Fukumura2017} of the order of the Roche lobe radius ($\sim 0.5a$). 
The mass loss in the \citet{Neilsen2016} model is highly (20-40 times) super-Eddington.
Our current abundance measurements suggest that perhaps the extended photosphere does not all need to originate in the disk, but can comprise in-situ ambient SN debris, which is heated by the X-ray source, thus alleviating the need for such high Eddington mass loss rates.

One of the unique features of the \gro\ outflow is its Fe$^{+21}$ absorption lines that enable density diagnostics \citep{Miller2006}. 
The density of a few $n \approx 10^{13}$\,cm$^{-3}$ \citep[][but see \citet{Tomaru2023}]{Miller2008, Mitrani2023} for an ion that forms at $\log \xi \approx 3$ indicates a distance from the ionizing source of a few $10^{10}$\,cm.
This ionization region is in the steep low-$\xi$ part of the AMD,  just below the break (Fig.\,\ref{fig:FullAMD}). Lower ionization (low-Z) gas would be more extended in the wind (as long as the density gradient is shallower than $n\sim r^{-2}$). 
It would therefore make sense that the SN contamination of the wind be in the outer regions, where the lower-$\xi$ (lower-$Z$, H-like) ions reside. 
Indeed, the abundances of elements with $Z \le 16$ seem to be under-represented by comparison the their theoretical SN yields (Figs.\,\ref{fig:SNmodelsZ}, \ref{fig:SNmodels}).

Can we say anything about the age of the SN remnant based on the accretion rate? The $6.6\,M_\odot$ BH in \gro\ grew either by a stellar merger, or by gradual mass accretion.
The Eddington mass accretion rate is $\dot m _{\rm acc} = L_{\rm Edd}/(\eta c^2)$, where $L_{\rm Edd}$ is the Eddington luminosity, $\eta \approx 10\%$ is the accretion efficiency, and $c$ is the speed of light. 
If the BH was formed at $5\,M_\odot$ \citep{Israelian1999, Gonzales2008}, for the current $6.6\,M_\odot$ BH, $L_{\rm Edd} \sim 10^{39}$\,erg\,s$^{-1}$ and  $\dot m _{\rm acc} \sim 10^{-7}\,M_\odot\,{\rm yr}^{-1}$. 
Hence, it would take the BH, accreting at the Eddington rate, $16$\,million years to gain $1.6\,M_\odot$, dating the SN to at least that far back.
However, the uncertainty on the initial BH mass, and the fact that accretion was not at the Eddington rate all the time, imposes large uncertainties on this age.

\section{Conclusions} \label{sec:conclusions}

This paper revisits the spectacular \hetg\ spectrum of the \gro\ 2005 outburst. Our analysis adds to the previous rich literature on this source several key results:

\begin{itemize}
    \item We report individual ionic column densities N$_{\rm ion}$ of 27 different ions (Table\,\ref{tab:ions}), pertaining to 18 different elements.

    \item We construct an absorption measure distribution ($AMD= dN_{\rm H}/d\log\xi$) that shows a sharp increase from low-$\xi$ that levels off at $\log \xi \approx 3.3$ (Fig.\,\ref{fig:FullAMD}).

    \item The AMD recovers the measured N$_{\rm ion}$ values and yields relative abundances of all 18 elements, which are fairly different from those of the binary companion (Table\,\ref{tab:abund}).

    \item The high abundances of odd-Z elements in the outflow are evidence for neutronization processes, which take place in high-mass ($\geq 25\,M_\odot$) high-metallicity ($\geq Z_\odot$) stars.
    

\end{itemize}

The measured abundances are thus a mix of a 25\,$M_\odot$ SN debris and the accretion disk material feeding the BH from its companion. This kind of analysis opens the door to more SN archaeology with X-ray binary outbursts.

%

\vspace{5mm}
\facilities{
}


\software{Xspec \citep{Arnaud1996}, XSTAR \citep{Bautista2001}
          }

\begin{acknowledgements}
We thank J. Neilsen for useful comments on the manuscript. We thank the referee for comments on the manuscript that helped to clarify the text. This work was supported by a Center of Excellence of The Israel Science Foundation (grant No. 2752/19). N.K. is supported by the Ramon scholarship sponsored by the Ministry of Science \& Technology, Israel. This paper employs a list of Chandra data sets,
obtained by the Chandra X-ray Observatory, contained
in doi:https://doi.org/10.25574/cdc.211.

\end{acknowledgements}

\bibliography{sample631}{}



\end{document}